# The massive black holes, high accretion rates, and non-tilted jet feedback, of jetted AGN triggered by secular processes

Chandra B. Singh[1] & David Garofalo[2]

[1]South-Western Institute for Astronomy Research (SWIFAR), Yunnan University, University Town, Chenggong, Kunming 650500, People's Republic of China
[2]Department of Physics, Kennesaw State University, Marietta, GA 30060, USA

## Abstract

That jetted active galactic nuclei (AGN) are also hosted in spiral galaxies is now well established. Our understanding of how such objects might fit in the radio loud AGN subclass has been described by Foschini and others over the past decade in that jets in spirals are weaker than those of radio galaxies and quasars because the black holes in spirals tend to be less massive. Recent data, however, may be pointing to a different picture which we describe. Unlike powerful jetted AGN in ellipticals, we illustrate from model perspectives, features of jets in spirals responsible for limiting both their power as well as their effect on their host galaxies. AGN triggered by secular processes fail to generate jet re-orientation, a key ingredient in the jetted AGN feedback mechanism in merger-triggered ellipticals that leads to the red-and-dead radio galaxies at low redshift such as M87. As a result, jetted AGN in spirals tend to live in a separate part of the parameter space compared to radio galaxies and quasars. Because of the absence of jet re-orientation and due to the relatively short-lived jet phases, jetted AGN in spirals are best compared to radio quiet or jetless AGN than any other jetted AGN subclass.

1. Introduction

For decades it appeared that the subset of AGN with powerful jets were hosted exclusively by elliptical galaxies, suggesting a merger origin to the jet formation phenomenon (Wilson & Colbert 1995; Sikora et al 2007). Over the last decade, however, the number of powerful jetted narrow line Seyfert 1 or NLS1 (so-called Γ-NLS1) discovered in spiral galaxies has progressively increased (e.g. Vietri et al 2022). These appear to be AGN hosted in spiral galaxies with narrow emission lines (Osterbrock & Pogge 1985) discovered to have jets detected in gamma rays (Komossa et al 2006; Yuan et al 2008; Abdo et al 2009; Calderone et al 2011). While a fraction of NLS1 appear to live in spiral galaxies, the morphology of the host galaxies of many NLS1 have not been determined. As a result, caution must be exercised in distinguishing between a class of objects that may not be entirely homogeneous from the host galaxy perspective. Apart from the presence of a jet, Γ-NLS1 are similar to NLS1 without jets, in that they are feeding their black holes at near Eddington rates (Foschini 2011). Just over a decade ago, Γ-NLS1 were estimated to have a range of black hole mass that failed to reach the billion or more solar masses measured in radio galaxies and quasars, and this was thought to be the reason for the relatively weaker jets from Γ-NLS1 compared to radio galaxies and quasars (Foschini 2011; Foschini 2014; Varglund et al 2022). Kiloparsec-scale jets similar to radio quasars but from spiral hosts have been found in recent years (Vietri et al 2022; Jarvela et al 2022) and the numbers of Γ-NLS1 has increased (Foschini et al 2021; 2022). Our interest is in Γ-NLS1 triggered by secular processes in spirals. Although the numbers are not large, Γ-NLS1 from disk-like galaxies with larger black hole mass have been discovered, yet their jet powers remain flat (Wu, Ho & Zhuang 2022). While the statistical significance may be weak, the host galaxies of these jetted AGN have been identified as disk-like. Hence, they deserve special focus. That jet powers in these disk-like Γ-NLS1 fail to scale with black hole mass as they appear to do in radio galaxies and quasars suggests the possibility that the black hole engine is different in the two subclasses of AGN. The fact that radio galaxies and quasars experience both near-Eddington as well as low accretion rates in very sub-Eddington accretion systems, unlike NLS1 which accrete at near-Eddington values, suggests that AGN feedback in NLS1 and Γ-NLS1, is different from that in radio galaxies and quasars. Unlike what has been assumed in the literature (Foschini et al 2015; Berton et al 2016; Paliya et al 2019) we will describe model prescriptions that do not associate NLS1 in spirals with flat spectrum radio quasars as their parent family. And the reason has to do with the triggering of the AGN by way of secular processes as opposed to mergers. The goal of our work is to describe this difference from theory.

In Section 2.1 we explore the data on jet powers, black hole mass, and accretion rates in Γ-NLS1 compared to radio galaxies and quasars and motivate the idea that the black hole engines and AGN feedback in Γ-NLS1 in spirals triggered by secular processes are different from those of radio galaxies and quasars. In Section 2.2, we attempt to understand this difference from a theoretical perspective. In Section 3 we conclude.

2. Discussion

2.1. Observations

Foschini (2011) showed that powerful jets from massive black holes at the centers of spiral galaxies in many so-called Γ-NLS1 is well established, which breaks the old paradigm that powerful jets can only emerge from ellipticals (see Foschini 2020 for a recent review). We reproduce the data of Foschini (2011) for jet power versus black hole mass for Γ-NLS1 representing jetted AGN in spirals and a sample of jetted AGN (flat spectrum radio quasars-FSRQ and BL Lac objects) from ellipticals in Figure 1. The black hole mass for the sample has been estimated using different techniques: for strong-line objects, the optical-UV bump is considered as a direct emission from the accretion disk, and maximum temperature is estimated which in turn gives an estimate of the black hole mass in the center while for lineless objects time scale for variability is compared with light crossing time to get an upper limit on black hole mass and the value of the magnetic field in the emitting region is considered for a lower limit on the black hole mass (details in Ghisellini et al. 2010 and references therein). Jet power comes from the radio luminosity at 15 GHz as a proxy and is determined from a fit including all the data. This likely skews jet powers for Γ-NLS1 to higher values so we do not compare these jet powers to those of other Γ-NLS1 shown in our other figures. Note how the purple Γ-NLS1 distribute themselves over a range of black hole mass that is lower than that for the FSRQ/BL LAC. The possibility presents itself that weaker jet powers for Γ-NLS1 in spirals is due to less massive black holes. Our next goal is to add more recently discovered jetted AGN in spirals with larger black hole mass than in Figure 1 and see whether jet power increases with black hole mass.

Wu, Ho and Zhuang (2022) have estimated jet powers for jetted AGN determined to be in spiral galaxies and plotted them as a function of black hole mass, comparing them to the jetted AGN of the Owen-Ledlow diagram. We reproduce their results in our Figure 2. In this case, the stellar mass of the host galaxy as well as radio luminosity is used for the black hole mass estimate. However, we only distinguish between jetted AGN in spirals (filled blue) versus those in ellipticals (empty black). We see that the blue objects of Figure 2 fill the region above 8.2 for the log of black hole mass, showing that even more massive black holes can be triggered in spirals. Despite this, we see a flatter jet power dependence on mass. For log of black hole mass less than 8.5 we calculate a slope for the data of 0.52 while for log of black hole mass above 8.5 the slope drops to 0.39. Although this data allows us to directly compare jet powers between disk galaxies and elliptical galaxies, the sample size is small, and the statistics therefore not significant. We therefore caution against using this as direct evidence of anything, but see this instead as an opportunity to connect to and describe theoretical ideas that indeed suggest this behavior. We argue that as more data becomes available, a new picture will emerge that is compatible with this preliminary evidence against the idea that the engine in jetted AGN hosted by spiral/disk galaxies is analogous to that in radio galaxies hosted by ellipticals. There is some engine-based difference that must be identified and that we describe in the next section.

In Figure 3 we plot the Γ-NLS1 of Paliya et al (2019), which also show that Γ-NLS1 have black hole masses up to order $10^9$ solar masses. To determine the black hole mass, they used single-epoch optical spectroscopy and considered emission properties of the standard thin disk with certain luminosity and inner and outer radii. While Γ-NLS1 have been found in spiral galaxies, a good fraction of Γ-NLS1 may form from minor mergers that tend to preserve the spiral nature of the host. Such objects would indeed be similar to radio galaxies formed in ellipticals and must therefore be eliminated from the sample. Because of the nature of the model, we must single out jetted AGN in disk-like galaxies that are not triggered by mergers, but by secular processes. Among the Γ-NLS1 of Paliya et al (2019), only one object shows strong signatures of disk-like structure, 1H 0323+342, shown in yellow. Given its relatively small black hole mass, such an object provides no additional support for our picture. Whether the remaining objects are found to be triggered by mergers or by secular processes is worth further exploration.

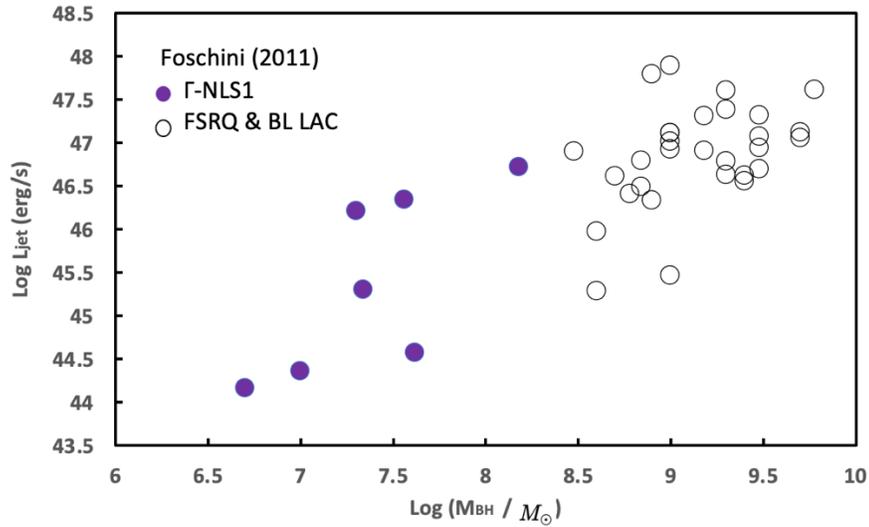

Figure 1: Γ-NLS1 and FSRQ/BL LAC objects from Foschini 2011, indicating the lower range of black hole mass and lower jet power for the former.

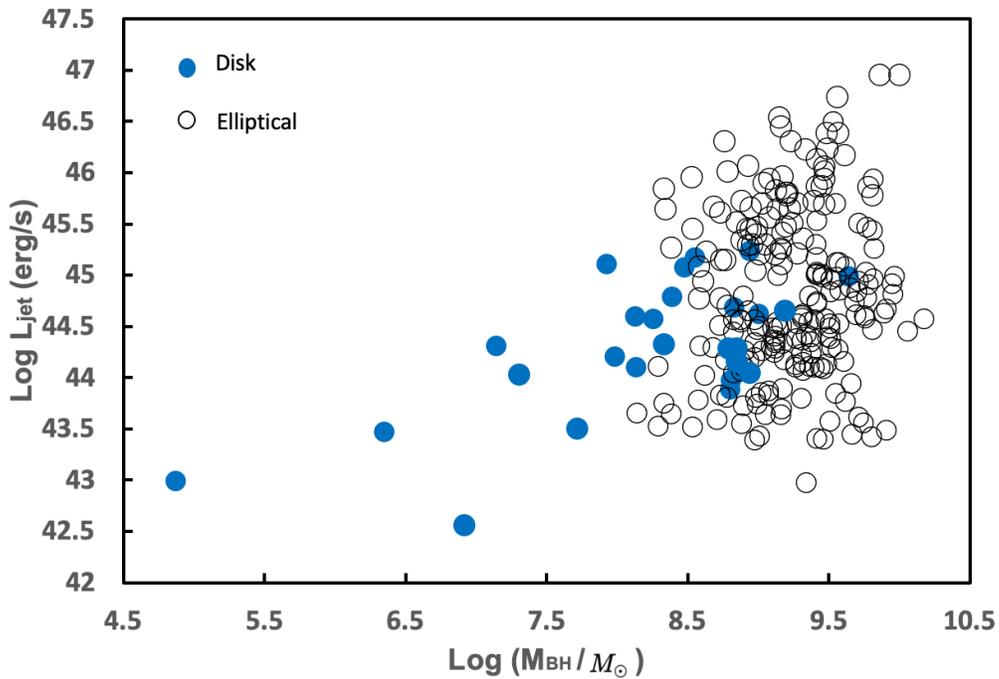

Figure 2: Jetted AGN from Wu, Ho & Zhuang (2022) with spiral/disk galaxies as filled blue and ellipticals as empty black showing disk galaxies with larger black hole mass compared to Figure 1, yet jet power remaining flat.

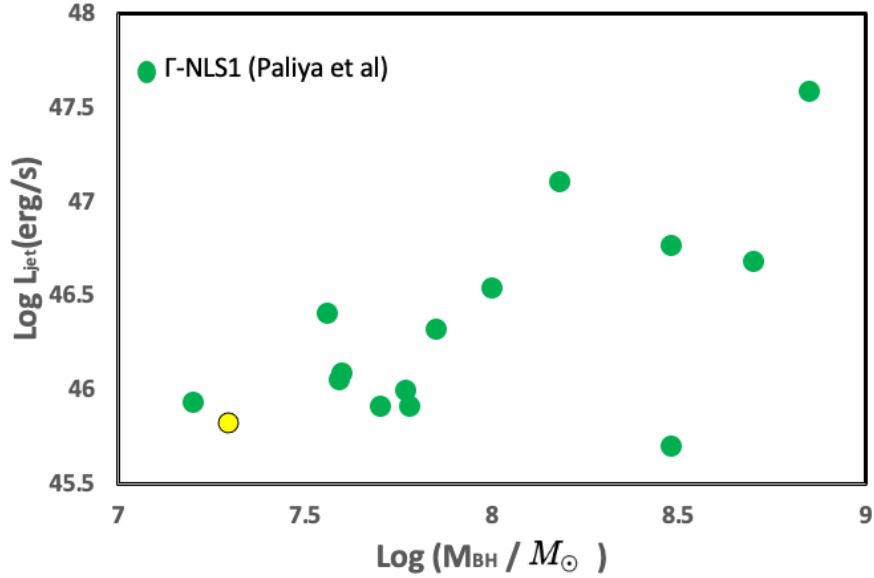

Figure 3: Γ-NLS1 from Paliya et al (2019) with one member of this class in yellow showing strong morphological features of disk-like galactic structure (Anton, Browne & Marcha 2008).

Finally, we plot the jet powers of a large sample of radio galaxies (from Poitevineau et al. 2023) along with the Γ-NLS1 of Paliya et al and of Foschini, versus their Eddington luminosities, λ = $L/L_{Edd}$, a proxy for accretion rates, with luminosity determined by the spectral energy distribution (see Ghisellini et al 2010). Besides taking data from available literature regarding the black hole mass estimate, Poitevineau et al. (2023) used a range of approaches like the relation between luminosity and FWHM under the assumption that broad line region is in virial equilibrium and single-epoch method like that of Paliya et al. (2019). The data reveals that while radio galaxies or jetted AGN triggered in ellipticals - likely from mergers - experience phases during which accretion rates are high as well as low, jetted AGN in disk/spiral galaxies only experience accretion rates that are compatible with radiatively efficient disks. This, we argue, constitutes another basic difference between jetted AGN in galaxies with different morphologies and triggering. In the next section, we address these differences from model perspectives.

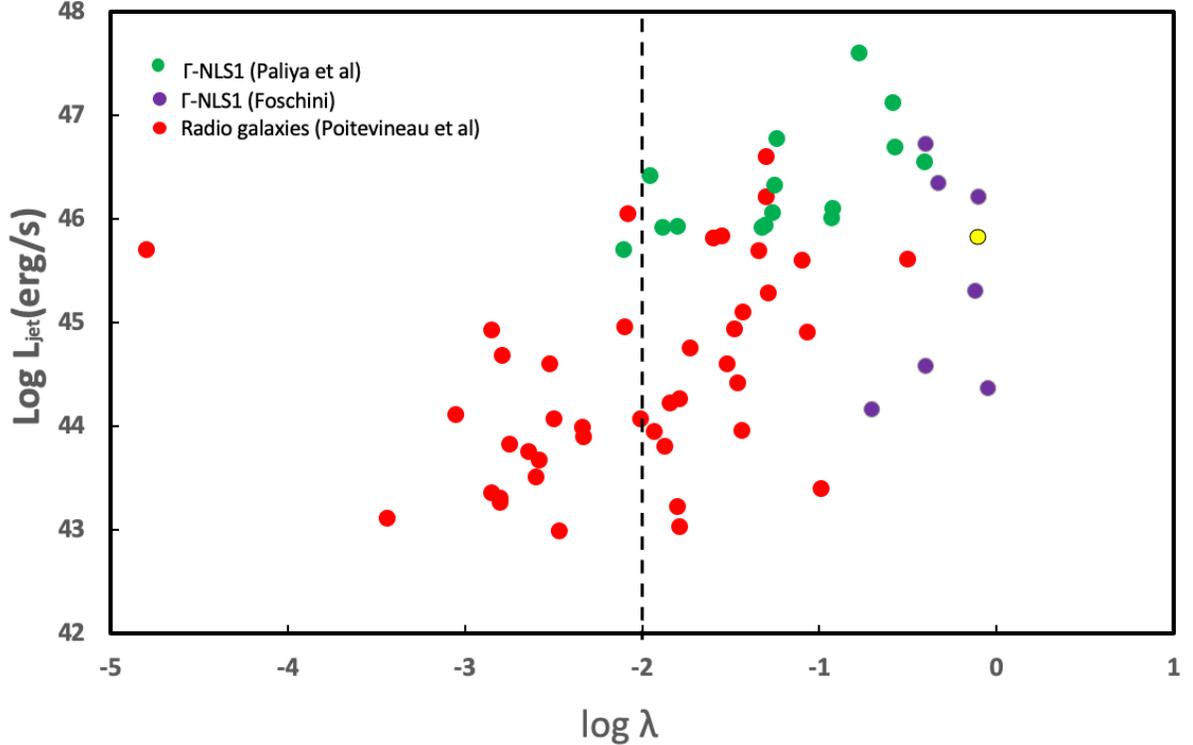

Figure 4: Eddington ratios ($\lambda = L/L_{Edd}$) indicating that Γ-NLS1, of which jetted AGN in spirals are a subclass, are overwhelmingly high accretion rate objects. 1H 0323+342 is again shown in yellow. The dashed line indicates the theoretical boundary between radiatively efficient accretion and advection dominated accretion.

2.2. Theory

Garofalo, Evans & Sambruna (2010) have produced a framework for understanding the triggering and evolution of AGN and the co-evolution of their host galaxies and their environment. In addition to combining the classic Blandford-Znajek and Blandford-Payne mechanisms to produce jets, as well as radiatively efficient and radiatively inefficient (ADAF) accretion, the model distinguishes itself with respect to the so-called spin paradigm by opening up the retrograde accretion window. Under restrictive conditions involving the angular momentum carried by the accreting material and the angular momentum of the black hole, counterrotation between accretion disk and black hole may form in a merger (King et al 2005; Garofalo, Christian & Jones 2019). Opening up the retrograde window for black hole accretion implies a time evolution that is not ad-hoc because counterrotation forces black holes to spin down and continued accretion will spin them up in the opposite direction. This fixed time evolution has allowed us to shed light on observations across the mass scale.

For our purposes, the focus is on AGN triggered in spiral galaxies (to model Γ-NLS1 formed in spirals and triggered by secular processes) and AGN triggered in mergers (to model radio galaxies and quasars). We restrict ourselves to illustrating the time evolution of two massive black holes that spin at 60% and 90% of their maximum rate, that are triggered in a spiral galaxy by secular processes (Figure 5), and in a merger in a rich environment (Figure 6), respectively. Part of the gas that is available to trigger the black hole also leads to abundant star formation, which is captured in the lowest panel and left columns of both Figures 5 and 6. The cold gas that triggers the black hole has settled into a thin, radiatively efficient accretion disk, captured by the thin blue line. Crucially, the Bardeen-Petterson effect (Bardeen & Petterson 1975) ensures that the disk settles into corotation with the black hole in the spiral galaxy as well as in most mergers in any environment, including rich ones. Hence, Figure 5 actually captures both the initial configuration for a black hole spinning at 60% its maximum rate triggered by both secular processes as well as those triggered by a merger. But such objects are subject to jet suppression (Neilsen & Lee 2009; Garofalo, Evans & Sambruna 2010; Ponti et al 2012) and do not therefore have jets for values of spin above about 0.7. The jet suppression phenomenon at high black hole spin in such systems allows us to appreciate why AGN triggered by secular processes struggle to form powerful jets even in systems with the most massive black holes (i.e. the black hole spin is never large).

A small parameter space exists for massive black holes such that counterrotation between black hole and accretion disk is stable (King et al 2005; Garofalo, Christian & Jones 2019), and the result is shown in Figure 6. The lowest panel of Figure 6, therefore, is different from the lowest panel of Figure 5 in that counterrotation forces the inner edge of the disk to live further away from the black hole. This ensures that jet suppression does not operate (Garofalo, Evans & Sambruna 2010; Garofalo & Singh 2016). Hence, a jet is generated.

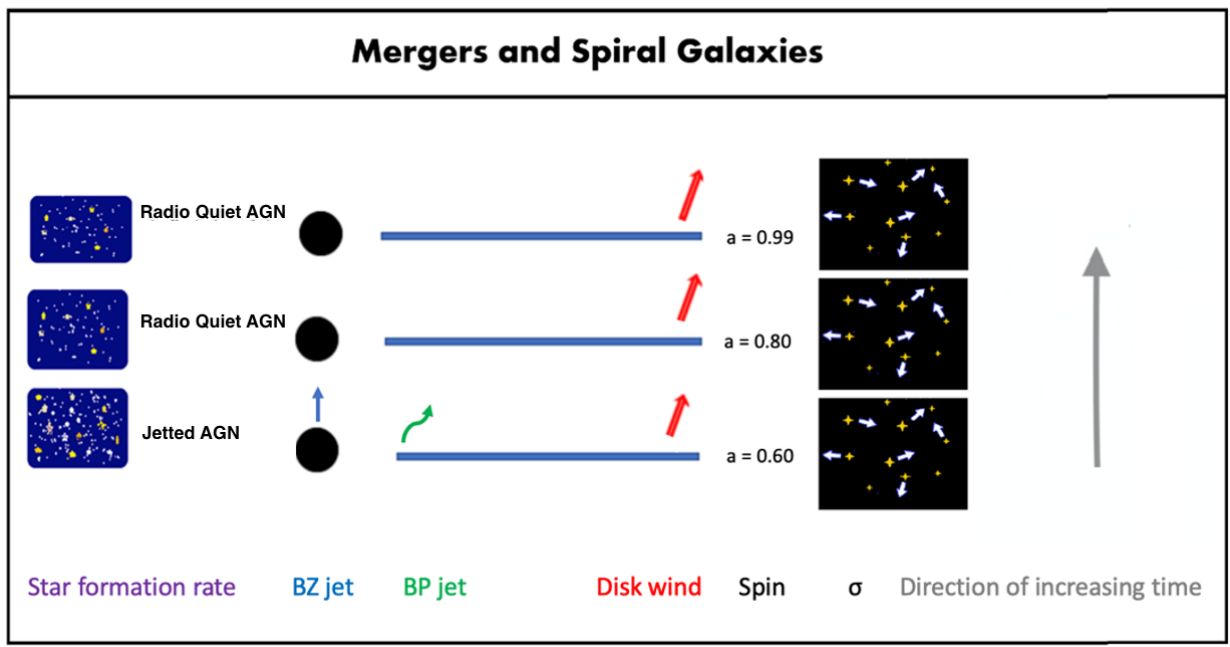

Figure 5: Spinning black hole triggered in a merger or by secular processes in a disk/spiral galaxy. The relatively weaker AGN feedback has little effect on the accretion rate. Left column: star formation rate; right column: Stellar velocity dispersion. BZ jet: Blandford-Znajek jet. BP jet: Blandford-Payne jet.

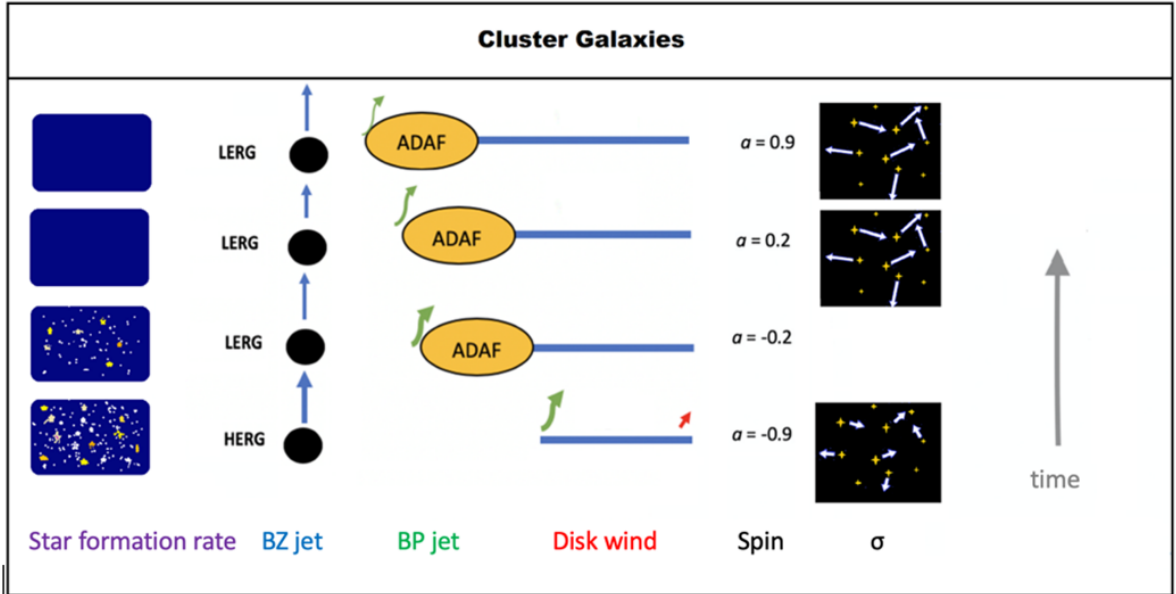

Figure 6: Black hole triggered in a merger in a rich environment. The strong AGN feedback affects the accretion disk and generates a transition from radiatively efficient accretion to advection dominated accretion (ADAF). The tilted jet in the transition through zero black hole spin is instrumental in suppressing star formation and enhancing stellar dispersion. BZ jet: Blandford-Znajek jet. BP jet: Blandford-Payne jet.

Accretion spins the black hole up in Figure 5 while it spins it down in Figure 6. The powerful jet generated in a near-Eddington accreting black hole (lowest panel of Figure 6) has a dominant feedback effect on accretion (Antonuccio-Delogu & Silk 2010; Garofalo, Evans & Sambruna 2010) as seen in the second panel from bottom of Figure 6. The accretion disk has evolved into an advection dominated disk (ADAF) in about 4 million years. The system becomes a low excitation radio galaxy, as the label LERG indicates. By contrast in the spiral galaxy, the absence of a jet precludes the feedback effect just described for the object illustrated in Figure 6. The object of Figure 5, therefore, simply spins the black hole up to higher values but no similar feedback effect from a jet is noticed in either the star formation rate or the stellar dispersion columns. Accretion is not the only victim of the jet feedback for Figure 6 objects. As the black hole spins down and then up in corotation with the accretion disk, the accretion disk likely experiences a tilt due to the absence of the Bardeen-Petterson effect at zero spin (Garofalo, Joshi et al 2019). This makes it so that when the spin is again sufficiently large to power a jet, the jet is tilted and this affects both star formation by shutting it down (the Roy Conjecture-Garofalo, Moravec et al 2022) as well as enhancing stellar velocity dispersion (Garofalo, Christian et al submitted). The absence of tilted jets in AGN triggered by secular processes is key to understanding why star formation rates tend not to be suppressed in such AGN compared to those in powerful low excitation radio galaxies (upper panels and left column of Figures 5 and 6).

Figures 5 and 6 reveal another crucial aspect of jets that we have yet to discuss, namely jet power. The paradigm is anchored to the idea that a powerful, collimated jet, is the result of both the Blandford-Znajek (BZ) and the Blandford-Payne (BP) mechanisms. The former involves magnetic fields anchored to the rotating black hole while the latter depends on magnetic fields threading the accretion disk. Both of these mechanisms are enhanced for greater magnetic flux threading the black hole (Garofalo, Evans & Sambruna 2010). And since greater magnetic flux threads the black hole for counterrotating configurations, the overall jet power for counterrotation is greater than for corotation (although see Tchekhovskoy & McKinney 2012). In short, for a given value of black hole spin, the counterrotating configuration produces a more powerful jet. From the theoretical expression for jet power given below (Garofalo, Evans & Sambruna 2010), we estimate the spin values of the black holes of the disk/spiral galaxies of the Wu, Ho & Zhuang (2022) sample.

$$L_{jet} = 2 \times 10^{47} \text{ erg s}^{-1} \, \alpha \beta^2 (B_d/10^5 G)^2 m_9^2 a^2 \qquad (1)$$

with

$$\alpha = 2.5 \, (1.5 - a) \qquad (2)$$

and

$$\beta = -1.5a^3 + 12a^2 - 10a + 7 - 0.002/(a - 0.65)^2 + 0.1/(a + 0.95) + 0.002/(a - 0.055)^2, \qquad (3)$$

with $a$ the spin parameter which is negative for counterrotation and positive for corotation and $B_d$ is the magnetic field threading the inner disk. We fix the magnetic field for all black holes with mass near $10^9$ solar masses and estimate the spin values for the spiral galaxies from their jet powers shown in Figure 2. We find their black holes to have spins between 0.1 and 0.7. It is also worth emphasizing that the difference in jet power between corotation and counterrotation is at two orders of magnitude for the extremes and that for the spin range of 0.1-0.7 it differs by only by about an order of magnitude. As a result, slight differences in accretion rates and/or magnetic field values can make this difference difficult to detect. The real power of our model framework is in the evolutionary history of the different systems, with radio galaxies in ellipticals experiencing the time evolution allotted to them by spin down followed by spin up. Overall, we have provided a theoretical explanation for both the potential observed relative weakness of jets in spiral galaxies despite black hole masses that overlap with those of more powerful jetted AGN in radio galaxies, and, more importantly, for the exclusively high accretion rates of jetted AGN triggered by secular processes in spirals compared to a range of accretion rates in radio galaxies, the latter being the tell-tale signature of black holes triggered by mergers into counterrotation.

3. Conclusions

Preliminary recent evidence points to jetted AGN in spirals possibly having black holes that can be as massive as those in radio galaxies and quasars, yet their jets tend to be weaker. In addition, jetted AGN triggered in spirals appear to be dominated by near-Eddington accretion rates, unlike their radio galaxy and quasar counterparts, which have both low accretion as well as high accretion rates. We have explained these properties theoretically from the perspective of a model which posits that counterrotation is key in the formation of both the most powerful jets as well as the most effective AGN feedback. Because our focus has been on jetted AGN in spirals, we have not explored low spinning black holes. But it is worth emphasizing that the model presented here, connects Γ-NLS1 in spirals to NLS1 in spirals, i.e. the jetted NLS1 to the non-jetted

NLS1 in spirals. If the black hole spin is near zero, the BZ mechanism fails, and we have a non-jetted NLS1. Once the spin reaches above about 0.1, we have a jetted NLS1 or Γ-NLS1 which remains jetted until the spin reaches about 0.7 (see also Garofalo & Singh 2022). Beyond that value of spin, jet suppression sets in and a non-jetted NLS1 is again instantiated. This evolutionary interchange between Γ-NLS1 and NLS1 may receive observational support from spin measurements, with the latter prescribed to occupy low and high spin values as opposed to intermediate spin values for Γ-NLS1.

## 4. Acknowledgments

CBS is supported by the National Natural Science Foundation of China under grant No. 12073021. We thank the anonymous referee for important insights that changed our work in a significant way.